\documentclass[english,PRA,twocolumn]{revtex4}
\usepackage[T1]{fontenc}
\usepackage[utf8]{luainputenc}
\usepackage{amssymb}
\usepackage{graphicx}
\usepackage{esint}
\usepackage{epstopdf}

\makeatletter
\@ifundefined{textcolor}{}
{%
 \definecolor{BLACK}{gray}{0}
 \definecolor{WHITE}{gray}{1}
 \definecolor{RED}{rgb}{1,0,0}
 \definecolor{GREEN}{rgb}{0,1,0}
 \definecolor{BLUE}{rgb}{0,0,1}
 \definecolor{CYAN}{cmyk}{1,0,0,0}
 \definecolor{MAGENTA}{cmyk}{0,1,0,0}
 \definecolor{YELLOW}{cmyk}{0,0,1,0}
}

\makeatother

\usepackage{babel}
\begin{document}

\title{Quantum Metrology with Dicke Squeezed States}

\author{Z. Zhang, L.-M. Duan}

\affiliation{Department of physics, University of Michigan, Ann Arbor, U.S.A.\\
and Center for Quantum Information, IIIS, Tsinghua University, Beijing, China}

\begin{abstract}
We introduce a new class of quantum many-particle entangled states,
called the Dicke squeezed (or DS) states, which can be used to improve
the precision in quantum metrology beyond the standard quantum limit.
We show that the enhancement in measurement precision is characterized
by a single experimentally detectable parameter, called the Dicke squeezing
$\xi_{D}$, which also bounds the entanglement depth for this class
of states. The measurement precision approaches the ultimate Heisenberg
limit as $\xi_{D}$ attains the minimum in an ideal Dicke state. Compared
with other entangled states, we show that the Dicke squeezed states
are more robust to decoherence and give better measurement precision
under typical experimental noise.
\end{abstract}
\maketitle
Precision measurement plays an important role for scientific and technological
applications. In many circumstances, precision measurement can be
reduced to detection of a small phase shift by use of optical or atomic
interferometry \cite{1,2,3}. The precision of the phase measurement
improves with increase of the number of particles in the interferometer. 
For $N$ particles in non-entangled (classical)
states, the phase sensitivity $\Delta\theta$ is constrained by the
standard quantum limit $\Delta\theta\sim1/\sqrt{N}$ from the shot
noise \cite{1,2,3}. Schemes have been proposed to improve the measurement
precision beyond the standard quantum limit by use of quantum entangled
states \cite{1,2,3,4,5,6}. Two classes of states are particularly
important for this scenario: the GHZ states \cite{5} (or called
the NOON states in the second quantization representation \cite{6})
and the spin squeezed states \cite{1,2,3}. A number of
intriguing experiments have been reported to prepare these states
and use them for quantum metrology \cite{7,8,9,10,11,12}. These states
are typically sensitive to decoherence and experimental noise \cite{13}.
As a result, the number of particles that one can prepare into the
GHZ state, or the maximal spin squeezing that one can achieve, are
both severely limited in experiments by noise.

In this paper, we introduce a new class of many-particle entangled
states for quantum metrology, which we name as the Dicke squeezed
(DS) states. The DS states have the following interesting features:
(i) They represent a wide class of entangled states that can be 
characterized by a single parameter
called the Dicke squeezing $\xi_{D}$ with $\xi_{D}<1$. The Dicke
squeezing parameter $\xi_{D}$ can be conveniently measured in experiments
from detection of the collective spin operator of $N$ particles.
It provides the figure of merit for application of the DS states in
quantum metrology: for states with $\xi_{D}$,
the phase sensitivity $\Delta\theta$ and the phase measurement precision
$d\theta$ both improve from the standard quantum limit $1/\sqrt{N}$
to the new scaling $\sqrt{\xi_{D}/N}$. The phase shift can be read
out through the Bayesian inference for the DS states. Under a fixed
particle number $N$, the parameter $\xi_{D}$ attains the minimum
$1/\left(N+2\right)$ under the ideal Dicke state, and the phase sensitivity
correspondingly approaches the Heisenberg limit $\Delta\theta\sim1/N$,
in agreement with the previous result on the Dicke state \cite{14,15}.
(ii) The entanglement of the DS states can also be characterized by
the squeezing parameter $\xi_{D}$. For a many-body system with particle 
number $N$, we would like to know how many particles
among them have been prepared into genuinely entangled states. This
number of particles with genuine entanglement is called the entanglement
depth for this system \cite{16,17}. A criterion proved in Ref. \cite{17}
by one of us indicates that $\xi_{D}^{-1}-2$ gives a lower bound
of the entanglement depth for any DS states with the squeezing parameter
$\xi_{D}$. (iii) Compared with the GHZ state or the spin squeezed
states, we show that the DS states characterized by $\xi_{D}$ are
more robust to decoherence and experimental noise such as particle
loss. Substantial Dicke squeezing $\xi_{D}$ remains under a significant
amount of noise under which spin squeezing would not be able to survive
at all.

For a system of $N$ particles, each of two internal states $a,b$
(with effective spin-$1/2$), we can define a Pauli matrix $\overrightarrow{\sigma_{i}}$
for each particle $i$ and the collective spin operator $\overrightarrow{J}$
as the summation $\overrightarrow{J}=\sum_{i}\overrightarrow{\sigma_{i}}/2$.
Note that the components of $\overrightarrow{J}$ can be measured
globally without the requirement of separate addressing of individual
particles. If the particles are indistinguishable like photons or
ultracold bosonic atoms, we can use the number of particles $n_{a},n_{b}$
in each mode $a,b$ to denote the states. In this notation (second
quantization representation), the GHZ state of $N$ spins $\left|aa\cdots a\right\rangle +\left|bb\cdots b\right\rangle $
(unnormalized) is represented by $\left|N0\right\rangle _{ab}+\left|0N\right\rangle _{ab}$
, the so called NOON state \cite{6}. In term of the mode operators
$a,\, b$, the collective spin operators have the form $J_{z}=(a^{\dagger}a-b^{\dagger}b)/2$,
$J_{x}=(a^{\dagger}b+b^{\dagger}a)/2$, $J_{y}=(a^{\dagger}b-b^{\dagger}a)/2i$.
A small phase shift $\theta$ can be measured through the Mach-Zehnder
type of interferometer illustrated in Fig. 1 by detecting the number difference of the output
modes (the output $J_{z}$ operator). The two beam splitters in the
interferometer exchange the $J_{z}$ and $J_{y}$ operators and the
phase shifter is represented by a unitary operator $e^{i\theta J_{z}}$
which transforms $J_{y}$ to $\cos\theta J_{y}-\sin\theta J_{x}$.
Assume the input state has mean $\left\langle \vec{J}\right\rangle =\left\langle J_{x}\right\rangle $
and minimum variance $\left\langle \left(\Delta J_{y}\right)^{2}\right\rangle $
along the $y$-direction. By measuring $\left\langle J'_{y}\right\rangle =\cos\theta\left\langle J{}_{y}\right\rangle -\sin\theta\left\langle J{}_{x}\right\rangle \approx-\theta\left\langle J{}_{x}\right\rangle $,
the phase sensitivity $\Delta\theta$ is characterized by $\sqrt{\left\langle \left(\Delta J_{y}\right)^{2}\right\rangle }/\left\langle J_{x}\right\rangle $.
This motivates definition of the spin squeezing parameter \cite{2,3}
\begin{equation}
\xi_{S}=\frac{N\left\langle \left(\Delta J_{y}\right)^{2}\right\rangle }{\left\langle J_{x}\right\rangle ^{2}}
\end{equation}
as the figure of merit for precision measurement. The phase sensitivity
is estimated by $\sqrt{\xi_{S}/N}$ for this measurement scheme.

Not all states useful for quantum metrology can be characterized by
the spin squeezing $\xi_{S}$. An example is the Dicke state $\left|N/2,N/2\right\rangle _{ab}$,
which has been shown to give the Heisenberg limited phase sensitivity
in \cite{14}. However, for this state $\left\langle \vec{J}\right\rangle =0$
in all the directions, and the spin squeezing $\xi_{S}$ is not a good
measure to characterize states of this kind with $\left\langle \vec{J}\right\rangle =0$.
To characterize a broad class of states that are useful for quantum
metrology, we introduce the following Dicke squeezing parameter, defined
as
\begin{equation}
\xi_{D}=\frac{N(\left\langle (\Delta J_{z})^{2}\right\rangle +\frac{1}{4})}{\langle J_{x}^{2}+J_{y}^{2}\rangle}.
\end{equation}
One can easily check that $\xi_{D}=1$ for the benchmark spin-coherent
states. We call any states with $\xi_{D}<1$ as the Dicke squeezed
states and a major result of this paper is to show that such states
are useful for quantum metrology where the phase sensitivity is improved
from $\sqrt{1/N}$ for the benchmark spin coherent state to about
$\sqrt{\xi_{D}/N}$ for the DS states. The parameter $\xi_{D}$ attains
the minimum $1/\left(N+2\right)$ under the ideal Dicke state $\left|N/2,N/2\right\rangle _{ab}$,
and the phase sensitivity $\sqrt{\xi_{D}/N}$ correspondingly approaches
the Heisenberg limit $\sim1/N$, in agreement with the result in \cite{14,15}.

The Dicke squeezing parameter $\xi_{D}$ also characterizes the entanglement
depth $E_{d}$ for many-particle systems. For an $N$-qubit system,
the entanglement depth $E_{d}$ measures how many qubits have been
prepared into genuinely entangled states \cite{16,17}. A theorem
proven in Ref. \cite{17} shows that $\left\lceil \xi_{D}^{-1}\right\rceil -2$,
where $\left\lceil \xi_{D}^{-1}\right\rceil $ denotes the minimum
integer no less than $\xi_{D}^{-1}$, gives a lower bound of the entanglement
depth $E_{d}$ . So the defined Dicke squeezing parameter $\xi_{D}$
provides a figure-of-merit both for entanglement characterization
and its application in quantum metrology, and this parameter can be
conveniently measured in experiments through detection of the collective
spin operator $\overrightarrow{J}$.

To show that $\xi_{D}$ is the figure-of-merit for quantum metrology,
we use two complementary methods to verify that the phase measurement
precision is improved to $\sqrt{\xi_{D}/N}$ for a variety of states. 
First, in the Mach-Zehnder (MS) interferometer
shown in Fig. 1a, the phase sensitivity is estimated by the intrinsic
uncertainty $\Delta\theta$ of the relative phase operator defined
between the two arms (modes $a_{\pm}$). We calculate this phase uncertainty
and find that it scales as $\sqrt{\xi_{D}/N}$ for various input states
with widely different $\xi_{D}$ and $N$. Second, we directly estimate
the phase shift $\theta$ by the Bayesian inference through detection
of the spin operator $J_{z}$, and find that the measurement precision,
quantified by the standard deviation $d\theta$ of the posterior phase distribution,
is well estimated by $\beta\sqrt{\xi_{D}/N}$ , where $\beta\approx1.7$
is a dimensionless prefactor. We perform numerical simulation of experiments
with randomly chosen phase shift $\theta$ and find that the difference
between the actual $\theta$ and the the measured value of $\theta$
obtained through the Bayesian inference is well bounded by 
$d\theta$, so $d\theta$ is indeed a good measure of the measurement
precision.

\begin{figure}
\includegraphics[width=6.5cm,height=2.5cm]{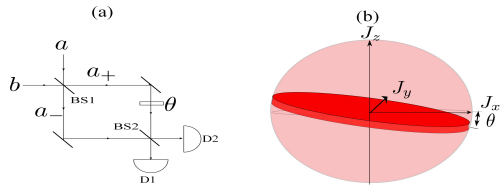}\caption{(a) The 
Mach-Zehnder (MZ) interferometer  to measure the relative
phase shift $\theta$ with input modes $a,b$ in Dicke squeezed states.
The detectors D1 and D2 measure the $J_{z}$ operator by recording
the particle number difference in the two output modes. (b) In the
Bloch sphere for the collective spin operator $\vec{J}$, a measurement
of the phase shift by the MZ interferometer is represented by rotation
of a thin disk (its size in $x,y,z$ directions corresponds to the
variance of $\vec{J}$ under the Dicke squeezed state) by an angle
$\theta$.}
\end{figure}

The Dicke state $\left|N/2,N/2\right\rangle _{ab}$ represents an
ideal limit, and it is hard to obtain a perfect Dicke state in experiments
in particular when the particle number $N$ is large. Here, we consider
two classes of more practical sates as examples to show that $\xi_{D}$
is the figure-of-merit for application in quantum metrology when the
ideal Dicke state is distorted by unavoidable experimental imperfection.
For the first class, we consider pure states of the form $\left|\Psi\left(\sigma\right)\right\rangle _{ab}=\sum_{n=0}^{N}a_{n}\left(\sigma\right)|n,N-n\rangle$,
where the total number of particles is fixed to be $N$ but the number
difference between the modes $a,b$ follows a Gaussian distribution
$a_{n}\left(\sigma\right)=exp\left\{ \frac{-(n-\frac{N}{2})^{2}}{\sigma^{2}}+i\frac{\pi}{4}(n-\frac{N}{2})\right\} $
with width characterized by $\sigma$. The
phase of $a_{n}\left(\sigma\right)$ is chosen for convenience so
that the variance of the state is symmetric along the $x,y$ axes.
For the second class, we consider mixed states $\rho_{ab}\left(\eta\right)$
which come from noise distortion of the Dicke state $\left|N/2,N/2\right\rangle _{ab}$
after a particle loss channel with varying loss rate $\eta$. To calculate
$\rho_{ab}\left(\eta\right)$, we note that a loss channel with loss
rate $\eta$ can be conveniently modeled by the transformation $a=\sqrt{1-\eta}a_{in}+\sqrt{\eta}a_{\nu}$
and $b=\sqrt{1-\eta}b_{in}+\sqrt{\eta}b_{\nu}$, where $a_{in},b$$_{in}$
denote the annihilation operators of the input modes which are in
the ideal Dicke state $\left|N/2,N/2\right\rangle =\left((N/2)!\right)^{-1}(a{}_{in}^{\dagger}b_{in}^{\dagger})^{N/2}|0,0\rangle$
and $a_{\nu}$,$b_{\nu}$ represent the corresponding vacuum modes.
By substituting $a{}_{in}^{\dagger},b_{in}^{\dagger}$ with $a{}^{\dagger},b^{\dagger}$
through the channel transformation and tracing out the vacuum modes
$a_{\nu}^{\dagger}$,$b_{\nu}^{\dagger}$, we get the matrix form
of $\rho_{ab}$$\left(\eta\right)$ in the Fock basis of the modes
$a,b$. The two classes of states $\left|\Psi\left(\sigma\right)\right\rangle _{ab}$
and $\rho_{ab}\left(\eta\right)$ approach the ideal Dicke state when
the parameters $\sigma,\eta$ tend to zero.

In the Mach-Zehnder interferometer shown in Fig. 1a, the modes $a_{\pm}$
of the two arms are connected with the input modes $a,b$ by the relation
$a_{\pm}=\left(\pm a+b\right)/\sqrt{2}$. The phase eigenstates $|\theta_{l}\rangle_{\pm}$
of the modes $a_{\pm}$ are superpositions of the corresponding Fock
states $|n\rangle_{\pm}$ with $|\theta_{l}\rangle_{\pm}=\left(s+1\right)^{-1/2}\sum_{n=0}^{s}e^{in\theta_{l}}|n\rangle_{\pm}$,
where $\theta_{l}=2\pi l/(s+1)$ ($l=0,1,...,s$) and $s+1$ denotes
the Hilbert space dimension which eventually takes the infinity limit
\cite{17a}. For modes $a_{\pm}$ in a composite state denoted by
its density matrix $\rho_{\pm}$, the probability distribution $P(\theta_{r})$
of the relative phase $\theta_{r}$ between the two interferometer
arms is given by

\begin{equation}
P(\theta_{r})=\sum_{l=0}^{s}\,_{\pm}\langle\theta_{l}\theta_{l-\delta l}|\rho_{\pm}|\theta_{l}\theta_{l-\delta l}\rangle_{\pm},
\end{equation}
where $\delta l=\theta_{r}(s+1)/(2\pi).$ The phase distribution $P(\theta_{r})$
becomes independent of the Hilbert space dimension $s+1$ when $s$
goes to infinity, and the standard deviation $\Delta\theta$ of $P(\theta_{r})$
gives an indicator of the intrinsic interferometer sensitivity to
measure the relative phase shift for the given input state \cite{14,15}.
We use $\Delta\theta$ to quantify the phase sensitivity for our input
states.

In Fig. 2, we show the calculated phase sensitivity $\Delta\theta$
for the two classes of input states $\left|\Psi\left(\sigma\right)\right\rangle _{ab}$
and $\rho_{ab}\left(\eta\right)$, by varying the parameters $\sigma,\eta$
and the particle number $N$. With fixed parameters $\sigma,\eta$,
when we vary the particle number $N$ (typically from $20$ to $200$
in our calculation), the phase sensitivity $\Delta\theta$ follows
a liner dependence with $\sqrt{\xi_{D}/N}$ by $\Delta\theta=\alpha\sqrt{\xi_{D}/N}$
(note that the Dicke squeezing parameter $\xi_{D}$ changes widely
as we vary $N$ and $\sigma,\eta$). The slope $\alpha$ depends very
weakly on the parameters $\sigma,\eta$ as shown in Fig. 2(c) and
2(d) and roughly we have $\alpha\approx2$. This shows that for different
types of input states the phase sensitivity $\Delta\theta$ is always
determined by the parameter $\sqrt{\xi_{D}/N}$ up to an almost constant
prefactor $\alpha$.

\begin{figure}
\includegraphics[width=8.5cm,height=4cm]{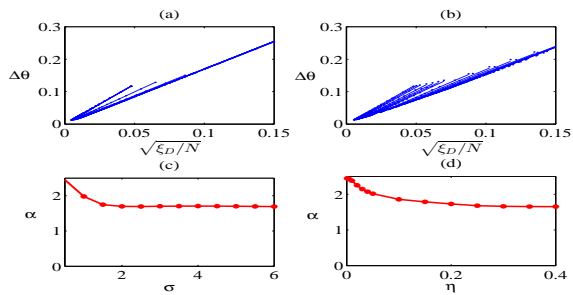}\caption{The phase sensitivity $\Delta\theta$ versus the normalized Dicke
squeezing $\sqrt{\xi_{D}/N}$ for two classes of input states: (a)
States $\left|\Psi\left(\sigma\right)\right\rangle _{ab}$ with Gaussian
superposition coefficients. (b) Dissipative states $\rho_{ab}\left(\eta\right)$
after a loss channel. The resulting points are on a straight line
when we vary the particle number $N$ from $20$ to $200$ ($\xi_{D}$
changes correspondingly) and the slope of the line changes slightly
as we vary the parameter $\sigma$ (from $0$ to $6$) or $\eta$
(from $0$ to $0.4$). (c) and (d) show the variation of the slope
$\alpha$ as a function of the parameter $\sigma$ or $\eta$. }
\end{figure}

A good phase sensitivity $\Delta\theta$ is an indicator of possibility
of high-precision measurement of the relative phase shift $\theta$,
however, the sensitivity by itself does not give the information of
$\theta$. In particular, for the DS sates we typically have $\left\langle \vec{J}\right\rangle =0$
and therefore cannot read out the information of $\theta$ by measuring
rotation of the mean value of $\vec{J}$. A powerful way to read out
the information of $\theta$ is through the Bayesian inference \cite{14,15}.
Here, we show that with the Bayesian inference, we can faithfully
extract the information of $\theta$ with a measurement precision
$d\theta=\beta\sqrt{\xi_{D}/N}$ for the DS states, where the prefactor
$\beta\approx1.7$. We note that each instance of measurement by the
MZ interferometer shown in Fig. 1 records one particular eigenvalue
$j_{z}$ of the $J_{z}$ operator, which occurs with a probability
distribution $P(j_{z}|\theta)$ (called the likelihood) that depends
on the relative phase shift $\theta$. With a given input state $\rho_{ab}$
for the modes $a,b$, the likelihood $P(j_{z}|\theta)$ is given by
.

\begin{equation}
P(j_{z}|\theta)=\langle j,j_{z}|e^{i\theta J_{y}}\rho_{ab}e^{-i\theta J_{y}}|j,j_{z}\rangle,
\end{equation}
where $|j,j_{z}\rangle$ denotes the momentum eigenstate with $j=N/2$.
The Bayesian inference is a way to use the Bayes' rule to infer the
posterior distribution $P_{m}(\theta|\left\{ j_{z}\right\} _{m})$
of the phase shift $\theta$ after $m$ instances of measurements
of the $J_{z}$ operator with the measurement outcomes $\left\{ j_{z}\right\} _{m}=j_{z1},\, j_{z2},\,\cdots,\, j_{zm}$,
respectively. After the $m$th measurement with outcome $j_{zm}$,
the phase distribution $P_{m}(\theta|\left\{ j_{z}\right\} _{m})$
is updated by the Bayes' rule
\begin{equation}
P_{m}(\theta|\left\{ j_{z}\right\} _{m})=\frac{P(j_{zm}|\theta)P_{m-1}(\theta|\left\{ j_{z}\right\} _{m-1})}{P(j_{zm}|\left\{ j_{z}\right\} _{m-1})},
\end{equation}
where $P(j_{zm}|\left\{ j_{z}\right\} _{m-1})=\int d\theta P(j_{zm}|\theta)P_{m-1}(\theta|\left\{ j_{z}\right\} _{m-1})$
is the probability to get the outcome $j_{zm}$ conditional on the
sequence $\left\{ j_{z}\right\} _{m-1}$ for the previous $m-1$ measurement
outcomes. Before the first measurement, the prior distribution $P_{0}(\theta)$
is assumed to be a uniform distribution between $0$ and $2\pi$.
When the rounds of measurements $m\gg1$, the posterior distribution
$P_{m}(\theta|\left\{ j_{z}\right\} _{m})$ is typically sharply peaked
around the actual phase shift, and we use the standard deviation $d\theta$
of $P_{m}(\theta|\left\{ j_{z}\right\} _{m})$ to quantify the measurement
precision.

To show that the measurement precision $d\theta$ is indeed determined
by $\sqrt{\xi_{D}/N}$ for the DS states, we numerically simulate
the MZ experiment with a randomly chosen actual phase shift $\theta_{r}$
in the interferometer. We take input states of the forms of $\left|\Psi\left(\sigma\right)\right\rangle _{ab}$
or $\rho_{ab}\left(\eta\right)$ as we specified before, with the
corresponding likelihood $P(j_{z}|\theta)$ given by Eq. (4). With
this likelihood, we get a sequence of measurement outcomes $j_{z1},\, j_{z2},\,\cdots,\, j_{zm}$,
which are sampled in our numerically simulated experiments using the
corresponding probability distributions $P(j_{zk}|\left\{ j_{z}\right\} _{k-1})$
with $k=1,\,2,\,\cdots,\, m$ , respectively. For this sequence of
outcomes, we obtain the corresponding sequence of posterior phase
distributions $P_{m}(\theta|\left\{ j_{z}\right\} _{m})$, with an
example shown in Fig. 3(a). One can see that the distribution $P_{m}(\theta|\left\{ j_{z}\right\} _{m})$
indeed gets increasingly sharper with $m$ and its peak approaches
the actual phase shift $\theta_{r}$. We use the the central peak
position $\theta_{p}$ of the distribution $P_{m}(\theta|\left\{ j_{z}\right\} _{m})$
as an estimator of the measured phase shift, and the difference $\theta_{pr}=\left|\theta_{p}-\theta_{r}\right|$
therefore quantifies the measurement error. This error $\theta_{pr}$
is typically bounded by $d\theta$, indicating there is no systematic
bias by this inference method.

\begin{figure}
\includegraphics[width=8.5cm,height=5cm]{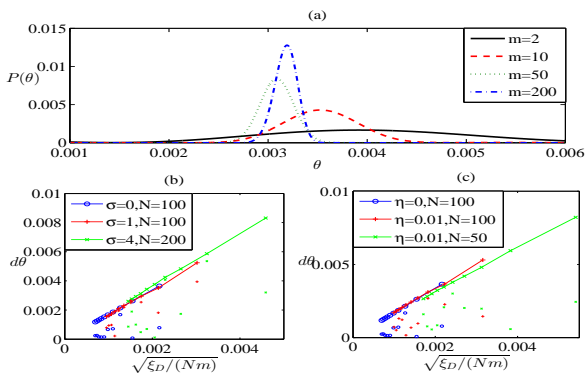}\caption{(a) The posterior phase distributions $P_{m}(\theta|\left\{ j_{z}\right\} _{m})$
obtained from the Bayesian inference after the $m$th measurement
with $m=2,10,50$ and $200$ from our numerically simulated experiments.
In the simulation, the actual phase shift $\theta_{r}=0.003$ and
the input state is $\left|\Psi\left(\sigma\right)\right\rangle _{ab}$
with $N=1000$, $\xi_{D}=0.0019$ and $\sigma=1$. (b) and (c): The
measurement precision $d\theta$ (the dots along a line fit by $d\theta\approx1.7\sqrt{\xi_{D}/\left(Nm\right)}$)
and the estimation error $\theta_{pr}$ (the scattered points below
the line) as functions of the scaled parameter $\sqrt{\xi_{D}/\left(Nm\right)}$
for the Gaussian input states $\left|\Psi\left(\sigma\right)\right\rangle _{ab}$
(b) and the dissipative input states $\rho_{ab}\left(\eta\right)$
(c) with $m$ varying from $20$ to $200$. The other parameters ($\sigma,N$
for $\left|\Psi\left(\sigma\right)\right\rangle _{ab}$ and $\eta,N$
for $\rho_{ab}\left(\eta\right)$) are specified by the inserts of
the figure. }
\end{figure}

In Fig. 3(b) and 3(c), we show the measurement precision $d\theta$
and the estimation error $\theta_{pr}$ as functions of the scaled
parameter $\sqrt{\xi_{D}/\left(Nm\right)}$, as we vary the types
of input states (the parameters $\sigma,\eta$ in states $\left|\Psi\left(\sigma\right)\right\rangle _{ab}$
and $\rho_{ab}\left(\eta\right)$), the particle number $N$, and
the rounds of measurement $m$. All the points for the measurement
precision $d\theta$ can be well fit with a linear function $d\theta\approx\beta\sqrt{\xi_{D}/\left(Nm\right)}$
with $\beta\approx1.7$. The estimation error $\theta_{pr}$ from
the simulated experiments (the scattered points) is typically below
the corresponding $d\theta$. This supports our central claim: the
defined Dicke squeezing parameter $\xi_{D}$ characterizes the improvement
of measurement precision for the DS states compared with the standard
quantum limit.

\begin{figure}
\includegraphics[width=8cm,height=2.5cm]{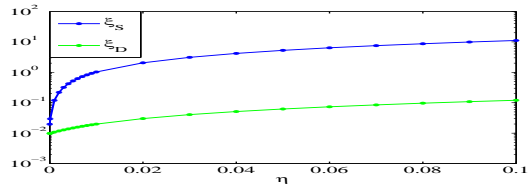}\caption{Comparison of the spin squeezing $\xi_{S}$ and the Dicke squeezing
$\xi_{D}$ under influence of the particle loss with a loss rate $\eta$.
We take the particle number $N=100$ and the amounts of squeezing
for $\xi_{S}$ and $\xi_{D}$ comparable initially at $\eta=0$. }
\end{figure}

Compared with other entangled states used in quantum metrology, a
remarkable advantage of the DS states characterized by the squeezing
parameter $\xi_{D}$ is its noise robustness. For instance, if the
noise in experiments is dominated by the dephasing error that does
not change the mode population, the numerator does not change in the
definition equation (2) for the Dicke squeezing $\xi_{D}$ and only
the denominator drops slowly. With a dephasing rate $p$ ($p$ is
the probability for each qubit to become completely decohered), the
squeezing parameter reduces to $\xi_{D}=1/\left[N\left(1-p\right)+2-p^{2}\right]$
if we start with a Dicke state for $N$ particles \cite{17}. We still
have substantial squeezing when $N\gg1$ even if the dephasing error
rate $p\gtrsim50\%$. More generic noise such as particle loss has
bigger influence on the Dicke squeezing, however, the DS states are
still more robust compared with other forms of entangled states such
as the spin squeezed states. In Fig. 4, we show the influence of the
particle loss to the Dicke squeezing $\xi_{D}$ and the spin squeezing
$\xi_{S}$, starting with comparable values of $\xi_{S}$ and $\xi_{D}$
at the loss rate $\eta=0$ under the same particle number $N$. The
spin squeezed state was determined by minimizing $(\Delta J_{z})^{2}$
with $J_{x}=0.1J$ \cite{16}. One can see that that spin squeezing
$\xi_{S}$ is quickly blown up by very small particle loss, but substantial
Dicke squeezing $\xi_{D}$ remains even under a significant loss rate.

In summary, we have proposed a new class of many-particle entangled
states, the DS states, to improve the measurement precision, and introduced the Dicke
squeezing parameter $\xi_{D}$ to characterize their performance in quantum metrology.
The Dicke squeezing is more robust compared with other forms
of entangled states. Substantial Dicke squeezing
can be generated in experiments, for instance, through the atomic
collision interaction in spinor condensates \cite{18,19}. With the
characterization and measurement method proposed in this paper, the
Dicke squeezing may find important applications for precision quantum
metrology.

This work was supported by the NBRPC (973 Program) 2011CBA00300 (2011CBA00302), the IARPA
MUSIQC program, the ARO and the AFOSR MURI programs.

\end{document}